# Heavy *d*-Electron Quasiparticle Interference and Real-space Electronic Structure of $Sr_3Ru_2O_7$

*Jinho Lee[1,2,3], *M. P. Allan[1,2], M. A. Wang[2], J.E. Farrell[1], S.A. Grigera[1,4], F. Baumberger[1], J.C. Davis[1,2,3] and A.P. Mackenzie[1]

[1] *Scottish Universities Physics Alliance, School of Physics and Astronomy, University of St. Andrews, St. Andrews, Fife KY16 9SS, Scotland.*

[2] *LASSP, Department of Physics, Cornell University, Ithaca NY 14853, USA.*

[3] *CMP&MS Department, Brookhaven National Laboratory, Upton, NY 11973, USA.*

[4] *Instituto de Fisica de Liquidos y Sistemas Biologicos, UNLP, La Plata 1900, Argentina.*

*These authors contributed equally to this work.

The intriguing idea that strongly interacting electrons can generate spatially inhomogeneous electronic liquid crystalline phases is over a decade old[1-5], but these systems still represent an unexplored frontier of condensed matter physics. One reason is that visualization of the many-body quantum states generated by the strong interactions, and of the resulting electronic phases, has not been achieved. Soft condensed matter physics was transformed by microscopies that allowed imaging of real-space structures and patterns. A candidate technique for obtaining equivalent data in the purely electronic systems is Spectroscopic Imaging Scanning Tunneling Microscopy (SI-STM). The core challenge is to detect the tenuous but 'heavy' k-space components of the many-body electronic state simultaneously with its r-space constituents. $Sr_3Ru_2O_7$ provides a particularly exciting **opportunity to address these issues. It possesses (i) a very strongly renormalized 'heavy' *d*-electron Fermi liquid[6,7] and (ii) exhibits a field-induced transition to an electronic liquid-crystalline phase[8,9] Finally, as a layered compound, it can be cleaved to present an excellent surface for SI-STM.**

The electronic structure of $Sr_3Ru_2O_7$ is somewhat complicated due to both its bilayer nature and the in-plane √2X√2 crystalline reconstruction caused by rotations of the $RuO_6$ octahedra (Fig. 1 inset). Nevertheless, photoemission studies reveal at least 5 Fermi surface pockets[6] in agreement with de Haas – van Alphen (dHvA) data[7]. The most striking result of these **k**-space spectroscopic measurements is that the intense electron-electron interactions cause bands that are 10 – 30 times flatter than they would be if quasi-free electrons existed in the system. Thus, although its metallic state is a Fermi liquid, $Sr_3Ru_2O_7$ is one of the most strongly renormalized heavy *d*-electron compounds known, and consequently, the spectral weight of **k**-space excitations should be dramatically reduced. This system undergoes a series of metamagnetic transitions in fields between 7.5 and 8.1 tesla applied parallel to the crystallographic c-axis. These transitions enclose an H-T plane region within which a large resistivity enhancement occurs[8]. Application of additional in-plane fields produces an 'easy' transport direction for which this enhanced resistivity disappears[10]. Thus, electronic transport with $180^O$ rotational ($C_2$) symmetry exists within this narrow H-T region while the surrounding regions display transport with the expected $90^O$ rotational ($C_4$) symmetry. These are the transport characteristics expected of a field-induced electronic nematic[10] - in this case, one generated from a many-body state of Ru *d*-electrons comprising both **r**-space and **k**-space spectral contributions.

Spectroscopic imaging scanning tunnelling microscopy (SI-STM) is potentially an ideal technique for studying such systems because simultaneous studies of **r**-space and **k**-space electronic structure can be performed. It was pioneered in studies of simple quasi-free electron systems[11,12] but, at least conceptually, might provide a route for imaging the far more complex many-body states of strongly correlated heavy-fermion systems. In SI-STM, measurement of the STM tip-sample differential tunnelling conductance $dI/dV(\mathbf{r},V) \equiv g(\mathbf{r},E=eV)$ at locations $r$ and sample-bias voltage $V$, yields an image proportional to the local density of electronic states LDOS(**r**,E). Moreover, **k**-space electronic structure elements can be determined simultaneously by using Fourier transform scanning tunnelling spectroscopy. This is because the spatial modulations in $g(\mathbf{r},V)$ due to interference of quasiparticles scattered by atomic-scale impurities is detectable in $g(\mathbf{q},V)$, the Fourier transform of $g(r,V)$. The **k**-space states undergoing the scattering interference have been deduced for both quasi-free electrons of noble metals[11,12] and the weakly renormalized electronic states of superconducting cuprates[13-18] in a process dubbed quasiparticle interference (QPI) imaging.

Multiple challenges exist in extending the basic techniques used for simple quasi-free electrons[11,12] to many-body states in a strongly correlated heavy fermion system like $Sr_3Ru_2O_7$. The larger the mass renormalisation due to electron-electron interactions becomes, the less the one-electron spectral weight in the **k**-space quasiparticles. Since SI-STM is a single electron spectroscopy, this loss of spectral weight is potentially a severe problem for detectability of **k**-space states by scattering interference. Another challenge is the complexity of the band structure[6,19]. In the case of simple metals and the cuprates, the reasons for the success of QPI is the combination of simple, un-renormalized band structures and a large joint density of states to enhance the intensity of the experimental signal[11-15]. But for multiple, topographically complex bands, one cannot assume that the equivalent process is even possible. Thus, detection of the heavy quasiparticles of a correlated many-body state using QPI appears extremely difficult, and, perhaps for this reason, it had not been attempted until now. Here, however, we are motivated by the possibility that if both the **r**-space and **k**-space electronic structure of $Sr_3Ru_2O_7$ can be imaged using techniques which may be applied in high magnetic fields, this would open the way for this material to become a 'hydrogen atom' for the study of the many-body states forming electronic liquid crystals.

To explore these issues we have carried out a SI-STM study on single crystals of $Sr_3Ru_2O_7$. The necessary scattering centers are introduced by replacing Ru atoms with Ti. In order to maximize the signal-to-background of QPI, we needed to introduce a sufficient number of scattering centers to generate the interference patterns but sufficiently few not to significantly alter the electronic structure. Accordingly, we grew single crystals of $Sr_3(Ru_{0.99}Ti_{0.01})_2O_7$, *i.e.* with 1% of the of Ru atoms replaced by Ti.

Our samples are cooled in cryogenic ultra-high vacuum, cleaved between adjoining SrO planes, and inserted in the STM head at or below 4.2K. Fig. 1a shows a 28 nm square topographic image of $Sr_3(Ru_{0.99}Ti_{0.01})_2O_7$. The quality of this image, and those previously obtained from $Sr_3Ru_2O_7$ [20] is related to a key advantage of Ruddlesden-Popper ruthenates for SI-STM studies of complex electronic matter. Each contains square planar $Ru^{4+}$ co-ordinated with four oxygen atoms in a basic $RuO_2$ formula unit. This means that, unlike in the cuprates, all planes in the material are charge-neutral, maximizing the possibility of obtaining cleaved surfaces that are bulk-representative. A top view schematic of the crystal structure is shown in the inset; it illustrates how Ti atoms substitute Ru in RuO$_2$ plane and how the √2X√2 reconstruction consists of sequential rotations of the RuO$_6$ octahedra about the c-axis. In fact, the topographic data in Fig. 1 reveal all these effects directly. Ti impurities from both upper and lower RuO layers are visible as dark and bright features respectively. The density of these features is in good agreement with the doped Ti density and they are absent in pure crystals. Moreover, the two different rotation angles corresponding to scattering from impurities in the two sublattices of each layer are also observable (labeled α and β for example in Fig. 1a).

In Fig. 1b we show the sub-unit-cell (*c.f.* Supplementary Information) differential conductance mapping g(**r**,E) measurements of a typical 5x5Å$^2$ field of view far from Ti impurities. Clear subatomic features are seen in the spatial arrangements of electronic structure, and there are several energies revealing **r**-space patterns expected of $d_{xz}, d_{yz}$ orbitals. The type of atomic resolution orbital occupancy imaging recently demonstrated in cuprates[21] may now become applicable to the Ru 4*d* states of Sr$_3$Ru$_2$O$_7$. The rate with which the **r**-space patterns evolve with energy is remarkable but could perhaps have been anticipated because, as we have emphasized, the electronic bands in Sr$_3$Ru$_2$O$_7$ are strongly renormalized by electron correlations[6,7]. In bands of mixed orbital character, the orbitals making the dominant contribution change as a function of **k** and energy. The images presented in Fig. 1b therefore probably represent the first real-space sub-atomic imaging of the energy evolution of strongly renormalized Wannier states; as such they invite more detailed theoretical investigation.

Figure 2A to 2F shows a sequence of 28nm square g(**r**,E) images taken simultaneously and focused on the filled states. The most striking features are the strong g(**r**,E) oscillations near

Ti atoms. Their dispersion is immediately obvious in the changes of g(**r**,E) in Fig. 2a to 2f. These changes occur because the LDOS(**r**,E) modulations exhibit characteristic wavevectors **q**(E) that disperse rapidly in energy. In Fig. 2g to 2l we show g(**q**,E), the Fourier transformed images of 2a to 2f respectively. The reciprocal-unit-cell locations of the Ru atoms are labeled as $(0,2\pi/a_0);(2\pi/a_0,0)$ where $a_0$ is the shortest inter Ru distance. Raw g(**q**,E) are presented in the supplementary Fig. 2. The dispersions of the QPI patterns are manifest by the rapidly changing g(**q**,E), as is the complexity of scattering processes. These data reveal, for the first time, scattering interference of very heavy *d*-electron quasiparticles at individual impurity atoms.

To identify the bands responsible for these interferences we consider the $Sr_3Ru_2O_7$ band structure[6,19] as outlined in Fig. 3a where it would seem that numerous scattering **q**-vectors are possible. But careful examination of the lengths, direction and dispersions of *all* the observed **q**(E) in Fig. 2 directs our focus to the $\alpha_2$ band which has primarily $4d_{xz},d_{yz}$ orbital character. Its **k**(E~$E_f$) contour is shown emphasized and overlaid on the LDA band structure[19] in Fig. 3a, and throughout the larger unreconstructed Brillouin zone in Fig. 3b. The $\alpha_2$-band spectral function A(**k**,E) is calculated from the E(**k**) estimated from ARPES (see supplementary materials) and the resulting regions of high joint-density-of states (JDOS) which dominate the QPI processes[14,16,17] as estimated by taking the autocorrelation[22,23] of A(**k**,E) are shown in Fig. 3c. Overlaid are a sequence of inequivalent scattering vectors $\mathbf{q}_i$:i=1…10 between the regions of high JDOS; these same vectors are shown in Fig. 3b to indicate how they interconnect the $\alpha_2$ band. Remarkably, all the inequivalent maxima in the complex g(**q**,E) between 5 meV and 13 meV below $E_F$ can be accounted for using this model. This is shown, for example in Fig. 3c, by overlaying the positions of these same $\mathbf{q}_i$ on g(**q**,E=-9meV) from the model as coloured arrows. The validity of the $\alpha_2$ QPI model is then borne out by the good agreement and highly overdetermined internal consistency between the model in Fig. 3b and data in Fig. 3d. Moreover, when the dispersions of the two basic vectors $\mathbf{q}_1$ and $\mathbf{q}_2$ (Fig 3b-d) are measured from Fig. 2g-l and plotted in Fig. 3e, they agree well with the directly measured dispersion of the $\alpha_2$ band from ARPES (see supplementary materials). In fact dispersions of the $\mathbf{q}_i(E)$ (which will be described elsewhere) are also in good agreement with these conclusions. Thus we find that the $\alpha_2$ band is detectable directly by FT-STS in $Sr_3Ru_2O_7$ and that it dominates QPI at Ti atoms. The velocities identified from the dispersion in Fig. 3e are approximately $1 \times 10^4$ ms$^{-1}$, in good agreement with estimates based on ARPES and dHvA[6,7]. These are ten times smaller than those seen in the $\alpha$ or $\beta$ bands of $Sr_2RuO_4$, confirming the high mass and strong correlations of the quasiparticles in $Sr_3Ru_2O_7$. As to why the $\alpha_2$ band should be predominant in QPI, one possibility is that strong momentum selective scattering effects due primarily to the correlations play a role[24].

These observations open the way for powerful modern SI-STM techniques[13-18] to be applied to many-body states not just in $Sr_3Ru_2O_7$, but also to a much wider class of strongly correlated heavy-fermion systems, including those that are spatially inhomogeneous. With regard to the problem of electronic nematicity in $Sr_3Ru_2O_7$, studies using the techniques introduced here but at high magnetic fields will be crucial. The current findings give at least two routes to microscopic investigation of a nematic state. First, the ability to resolve sub-atomic structure representative of the underlying orbitals offers the possibility of direct observation of lowered symmetry in their arrangement within the nematic region. Second, it is conceivable that the heavy *d*-electron QPI response will also show lowered symmetry. The density of Ti atoms chosen for the work reported here increases the residual resistivity of $Sr_3Ru_2O_7$ to ~10 μΩcm.

This does not cause either an electronic or structural phase transition or a change to the Fermi energy, but would wash out the bulk signals of nematicity. However, the Ti level in the crystals can be controlled with high precision, and now that the basic characteristics of heavy *d*-electron QPI have been established, it will be possible to study samples with an order of magnitude fewer Ti inclusions.

The observations that the strongest QPI signals come from the $\alpha_2$ band, and that **r**-space structures associated with the same $d_{xy}, d_{yz}$ orbitals are detectable, are both potentially important. Early theoretical studies of the metamagnetism in $Sr_3Ru_2O_7$ focused on a proposed van Hove singularity and explained metamagnetism as due to spin splitting of its high density of states[24] After the discovery of coexisting nematicity[10], a heuristic quadrupole electron-electron interaction connecting metamagnetism and nematicity was proposed[26]. More recent theoretical advances propose spatially heterogeneous states caused by these singularities[27], or model the fundamental microscopic cause of metamagnetic-nematicity in $Sr_3Ru_2O_7$ as due to the natural uni-directionality of underlying $d_{xz}$ $d_{yz}$ orbitals and the related quasi-one dimensional van Hove singularity[28,29]. Thus, in addition to our results having broad significance to the study of strongly renormalized metals, both the **r**-space and **k**-space electronic structure we report here may have relevance to the formation of an intriguing electronic liquid crystal phase.

Correspondence and requests for materials should be addressed to A.P.M.

**Acknowledgments** We acknowledge and thank E. Fradkin, T. Hanaguri, C.A. Hooley, E.-A. Kim, S. A. Kivelson, Y. Kohsaka, M. Lawler, A.J. Millis, S. Raghu, T.M. Rice, S. Sachdev, K.

Shen, H. Takagi, A. Tamai and F.-C. Zhang for helpful discussions and communications. These studies are carried out with support from NSF DMR-0520404 to the Cornell Center for Materials Research, from Brookhaven National Laboratory and from the UK EPSRC, Royal Society and Leverhulme Trust.

**Author contributions** The crystals were grown by J.E.F., SI-STM experiments performed by J.L., M.P.A. and M.A.W, and project planning, data analysis and paper-writing by J.L., M.P.A., S.A.G., F.B., J.C.D. and A.P.M.

# Figures

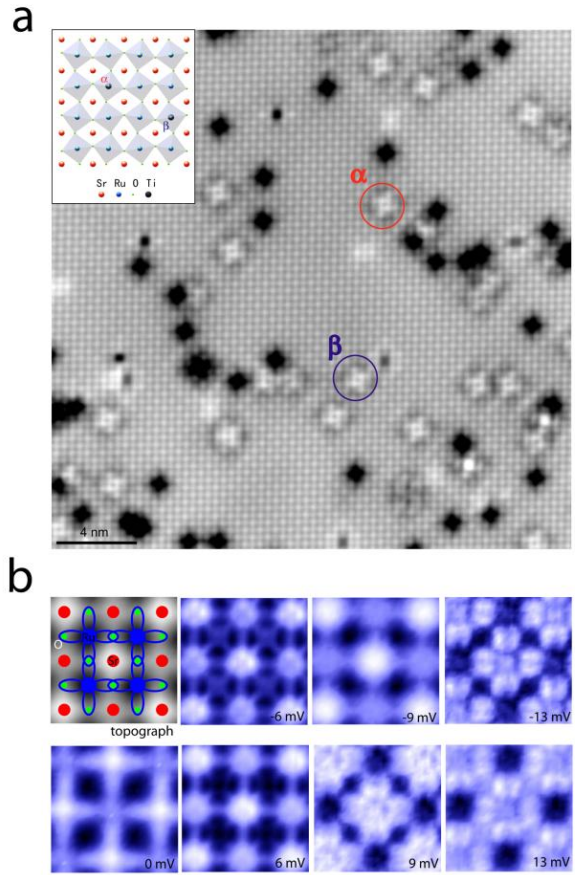

**Figure 1 Topography and sub-unit-cell electronic structure imaging** in **Sr$_3$Ru$_2$O$_7$**

**a** Topographic image of the SrO cleaved Sr$_3$Ru$_2$O$_7$ surface, taken at V = -100mV and 10GΩ. The top inset shows a schematic view from above along the c-axis showing the sequential 6.8$^o$ rotations of the RuO$_6$ octahedra which double the unit cell. The Ti dopant site is shown in black and two types of octahedra are labeled α and β. On the topographic image, dark and light spots stem from Ti impurities located at the Ru sites on the higher and lower Ru-O sheet of the top bilayer respectively. The white Ti-sites appear in two different orientations, corresponding to the different RuO octahedra orientations (inset). Data for Fig 1b were taken far from any Ti dopant atom.

**b** The top left-hand panel shows the locations of Ru atoms and their $d_{xz}$ and $d_{yz}$ orbitals in blue. Red and green circles mark the positions of Sr, and O atoms, respectively. Each subsequent panel shows g(**r**,E) maps resolving sub-unit-cell spatial features in the same field of view. While some g(**r**,E) show high intensity mainly at the positions of the Sr atoms (-9, 0 meV), others clearly resolve sub-unit-cell features with the symmetry and location of the $d_{xz}$, $d_{yz}$ orbitals (-13, +9, +13 meV).

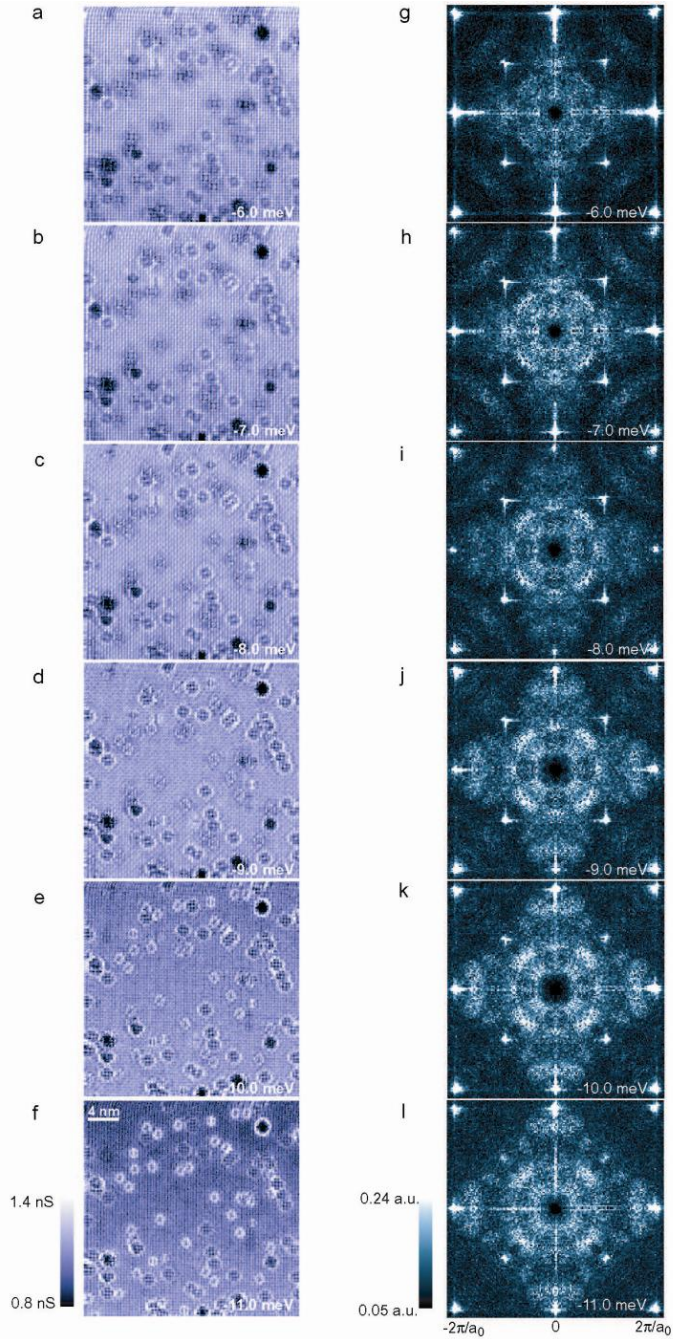

**Figure 2**      **Fourier Transform STS of $Sr_3Ru_2O_7$**

**a** A sequence of g(**r**,E) maps taken at V= -100mV, 1GΩ in the same 28nm-square field of view. Each Ti scatterer exhibits energy-dispersive QPI fringes around it.

**b,** The corresponding two dimensional Fourier transform image g(**q**,E) revealing heavy *d*-electron QPI directly. The dark area near (0,0) is where spectral weight has been reduced to allow for clearer viewing of the g(**q**,E) contrast and the images are octet-symmetrized (see Supplementary Information). A complex and fast-dispersing set of wavevectors $q_i$ is seen in these g(**q**,E). Remarkably, this **q**-space complexity and dispersion can be explained by scattering between states in only one very simple band of $Sr_3Ru_2O_7$.

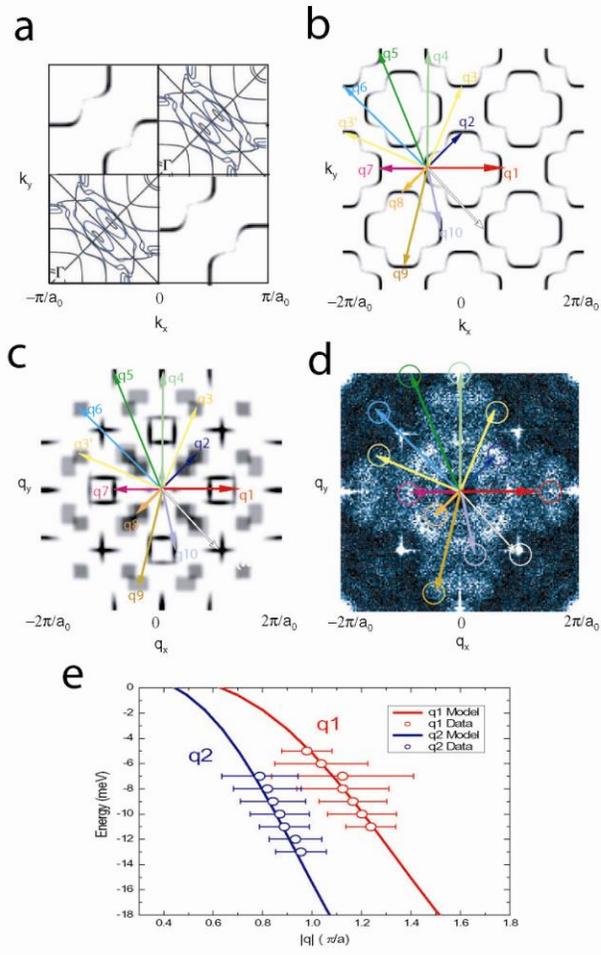

**Figure 3      Quasiparticle Interference in the $\alpha_2$ band of $Sr_3Ru_2O_7$**

**a,** The LDA(local-density-approximation) Fermi surface of $Sr_3Ru_2O_7$ in the first Brillouin zone with the estimated $\alpha_2$ band A($\mathbf{k}$,$E\sim E_F$).

**b,** The model A($\mathbf{k}$,$E$) of the $\alpha_2$ band in the extended zone scheme of the $\sqrt{2}\times\sqrt{2}$ reconstructed system at $E$ = -9 meV. The full set of inequivalent scattering vectors $\mathbf{q}_i$ ;$i$=1,…10 involving only the $\alpha_2$ band is shown as colored arrows (same arrows in **c,d**). This A(**k**,*E*) is calculated from the $\alpha_2$ band model estimated from ARPES measurements.

**c,** The autocorrelation of the model A(**k**,*E*) shown in **b.** This process picks out the regions of high JDOS which should dominate the quasiparticle scattering process. The full set of inequivalent scattering vectors **q**$_i$ ;*i*=1,…10 which should exist for scattering interference only in the $\alpha_2$ band is determined from these regions of high JDOS and is shown using the set of colored arrows.

**d**, By overlaying as open circles the tip positions of these same **q**$_i$ on g(**q**,*E*=-9 meV) we see that all inequivalent maxima can be accounted for with highly overdetermined internal consistency by $\alpha_2$ band scattering interference.

**e**, Measured dispersions of **q**$_1$ and **q**$_2$ from data in Fig. 2g-l. They agree well with the model $\alpha_2$ band underlaid as solid line and thus with the directly measured dispersion of the $\alpha_2$ band from ARPES (see Supplementary Information). Error bars indicate standard deviation widths of **q**$_1$ and **q**$_2$ peaks along the dispersion lines after fitting them to Gaussian curves at each energy.

**Methods**

*Crystal growth*

The crystals used for this work were grown in St Andrews in a NEC Machinery Corporation Image furnace (SCI-MDH-11020). Batches of $Sr_3(Ru_{1-x}Ti_x)_2O_7$ were prepared by mixing $RuO_2$, $TiO_2$ and dried $SrCO_3$ powders in the ratio (1-x)2.52: 2x: 3. The mixed powder was then reacted in air at 1200 ºC for 12 hours, reground, formed into a feed-rod and sintered on a layer of strontium ruthenate bedding powder for 2 hours at 1420 ºC, again in air. The rod was then melted in the image furnace, with crystalline growth seeded from the molten zone using a ruthenate seed crystal, and the polycrystalline rod fed into the molten zone at 25mm/hr. Gross phase purity was established using x-ray diffraction of powdered segments of the resultant single crystals and electron probe microanalysis (EPMA) was employed to determine the Ti concentration. The residual resistivity was measured using standard four-probe techniques and an a.c. method. Typical currents were 100 µA, and frequencies in the range 50 – 100 Hz. Data were taken in three cryostats: a continuous flow $^4$He cryotstat with a base temperature of 3.8K, an adiabatic demagnetisation refrigerator (Cambridge Magnetic Refrigeration) with a base temperature of 100 mK and a $^3$He-$^4$He dilution refrigerator (Oxford Instruments) with a base temperature of approximately 30 mK.

Magnetic characterisation was used to measure and then minimise the levels of $SrRuO_3$, $Sr_2RuO_4$ and $Sr_4Ru_3O_{10}$ impurity phases which are always a challenge when growing this member of the Ruddlesden-Popper series of layered perovskite ruthenates. $Sr_2RuO_4$ inclusions can be identified by studying the diamagnetic a.c. response on entering the superconducting state. For quantitative work, care must be taken to operate in the low demagnetising factor configuration in which the a.c. field is applied parallel to the *ab* planes. $SrRuO_3$ and $Sr_4Ru_3O_{10}$ are ferromagnetic with different Curie temperatures, so d.c. magnetisation measurements in a commercial SQUID magnetometer (Quantum Design MPMS) allows a sensitive determination of. small impurity contents. Full details of the characterisation procedures can be found in Ref. 30.

*Fourier Transform STS Measurements*

To obtain QPI in **q** space, we calculate the absolute intensity of the 2D FFT (2 dimensional fast Fourier transform) of the conductivity map g(**r**, E). This shows an extremely bright spot in the center due to long wavelength fluctuations. To remove these very small **q** vector elements in q-space, we first subtract the data value averaged with a Gaussian filter width of about 6 Å from the original conductance value at every pixel. The 2D Fourier transform of this resulting value effectively gets rid of the low **q** vector components, showing clearer individual QPI features.

Due to drift during a scan, the raw g(**r**, E) does not show perfect square symmetry, but is slightly elongated in one direction. To correct this effect, we use a conformal mapping algorithm to make g(**r**, E) a square. Here we use information about the atomic Bragg peak positions to calculate the necessary parameters of the conformal transformation. Although the resulting Fourier transform g(**q**, E) shows clear dispersive QPI features with 4 fold symmetry, we attempted to enhance the signal to noise ratio even further. We selectively averaged symmetrical data points over 2 octants of g(**q**, E) showing more discernible intensity peaks which, in this study, happened to be along the vertical direction. Mathematically this operation corresponds to symmetrically averaging (averaging only among symmetrically equivalent points under 4 fold symmetry) over the q- space region which satisfies $|q_y| > |q_x|$ and using only this part to construct complete g(**q**, E).

[30] Farrell, J.E. *The influence of cation doping on the electronic properties of $Sr_3Ru_2O_7$*. PhD thesis, University of St Andrews, on-line at http://hdl.handle.net/10023/689

# Supplementary Information

## 1. Supplementary Methods

*Unit Cell Averaging*

The $Sr_3Ru_2O_7$ surface forms square Bravais lattice, i.e. every point $(x_i, y_i)$ is equivalent to any other point $(x_i + n \cdot \sqrt{2}\, a_0, y_i + m \cdot \sqrt{2} a_0)$ where $n,m$ are integers, $a_0$ is the shortest inter Ru distance and the $\sqrt{2}$ factor accounts for in-plane superstructure induced by the rotations of the $RuO_6$ octahedra.

Given our ability to image large, flat area of the $Sr_3Ru_2O_7$, we can use the translational invariance to increase the signal-to-noise ratio (S/N) by averaging over many of these Bravais lattices.

First, we have to pinpoint the center of each unit cell. For this purpose, we use the simultaneously taken topographic image where Sr atoms appear as peaks. We determine the position of each Sr atom with an accuracy of less than 0.2 Å by fitting topographic peaks to a two dimensional Gaussian function. This way, we can precisely locate the position of every unit cell of the Ru-Ru lattice. To include the $\sqrt{2}$ superstructure of the lattice, we only use every other Ru atom. Second, we crop a ~15×15 Å$^2$ window around every unit cell center. We now have a set of maps which are ~15×15 Å$^2$ in size, centered about a crystallographically equivalent site.

Finally, we add all the equivalent points in these maps together, and divide the resulting sum by the number of maps, to arrive at a unit-cell averaged map with enhanced S/N.

The sub-atomic resolution maps presented in Fig. 1b are the result of averaging over 28 such unit cells.

## 2. Supplementary Discussion

*Agreement between the Model $\alpha_2$ Band and Photoemission Experiments*

The $\alpha_2$ band topology described in this paper has been constructed primarily to model the QPI data. To this end, we started from two strongly unidirectional bands $E_x(\mathbf{k})$, $E_y(\mathbf{k})$ described by the empirical formula[27]

$$E_{x,y}(k_x,k_y) = [1222 \cdot \sin^6(k_{x,y}a_0/2) - 26.1 \cdot \sin^4(k_{y,x}a_0/2) + 1.37] \text{ (meV)},$$

where $a_0$ is the Ru-Ru nearest neighbor distance. $E_x(\mathbf{k})$, and $E_y(\mathbf{k})$ corresponds to the bands of ruthenium $d_{xz}$ and $d_{yz}$ orbital character respectively. The $\alpha_2$ band is then obtained by hybridizing these hypothetical bands:

$E_{\alpha 2}(k_x,k_y) = (E_x + E_y)/2 + [(E_x - E_y)^2/4 + V^2]^{1/2}$ near highly degenerate direction($\Gamma$M) with $V$ = 2 meV. Apart from the points where two bands overlap, we can choose the band with higher energy of two($E_{\alpha 2}(\mathbf{k}) = \max(E_x(\mathbf{k}), E_y(\mathbf{k}))$) to construct $\alpha_2$ band[27]. Similar method can be also used to construct $\alpha_1$ band.

In Supplementary Fig. 2a, the model Fermi surface is overlaid on the experimental ARPES data[8]. Clearly, its size and shape closely resemble the cross-shaped experimental Fermi surface contour of $\alpha_2$ band derived from the out of plane $d_{xz}$, $d_{yz}$ orbitals[7,24]. The best agreement is observed after shifting the model band by 3 meV towards higher energies. This shift is probably due to small energy uncertainty as well as momentum uncertainty of both techniques, not due to possible different doping level since Ti impurities do not introduce change of $E_F$ in $Sr_3(Ru_{1-x}Ti_x)_2O_7$. The agreement between the model and ARPES extends to other energies: the dispersion plot along ΓM (Supplementary Fig. 2b) shows that the model $\alpha_2$ band has almost exactly the same Fermi velocity as the corresponding band measured by ARPES. Hence our QPI result is in very good agreement with the band structure measured by ARPES over the entire energy range investigated in this paper.

# 3. Supplementary Figures

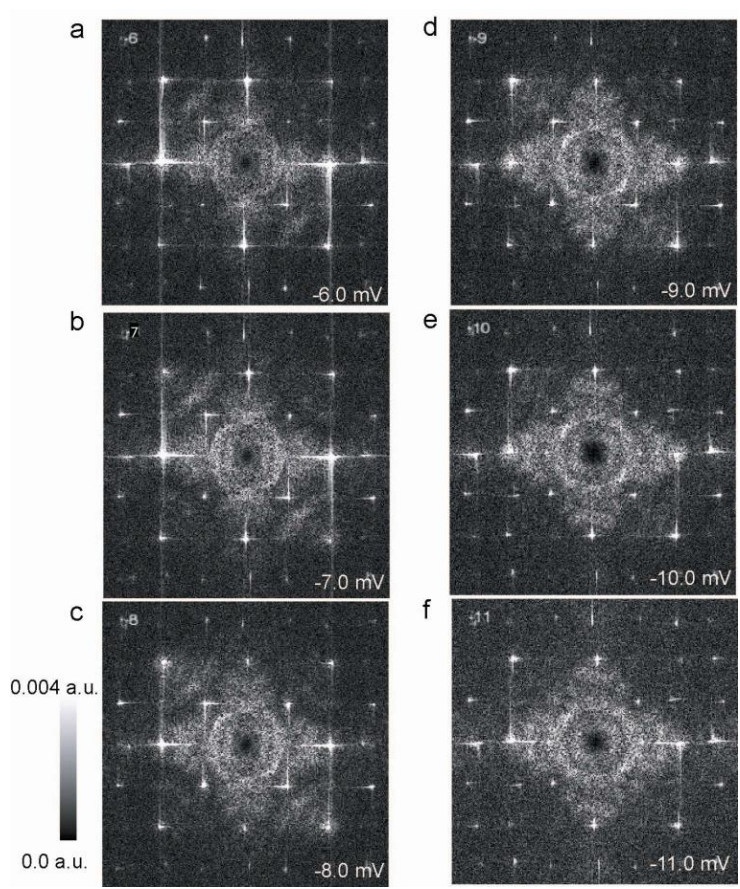

**Supplementary Figure 1 a-f**, 2D-FFT of the conductance value subtracted by long wave length averaged values at energies ranging from -6 meV to -11 meV. Highly intense central part which corresponds to the long wave length components has been effectively removed and now it appears as dark spot.

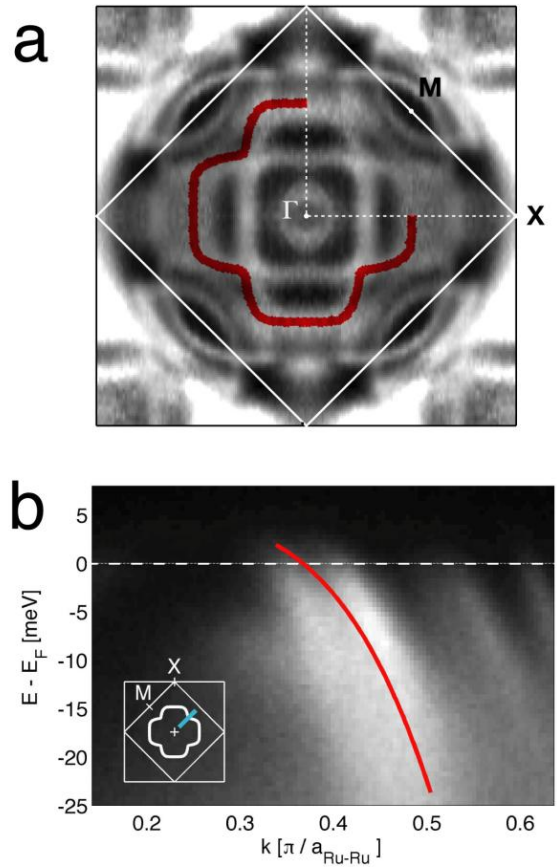

**Supplementary Figure 2** Agreement between the model $\alpha_2$ band and photoemission data. **a**, Comparison of Fermi surface (reproduced from ref. 8). The red contour stems from the model $\alpha_2$ band. **b**, Comparison of band dispersion along ΓM. The red line from the model $\alpha_2$ band follows regions of high photocurrent intensity.